# Cost-Performance Trade-off in Thermoelectric Air Conditioning System with Graded and Constant Material Properties

Abhishek Saini, Sarah J. Watzman, and Je-Hyeong Bahk*

*Department of Mechanical and Materials Engineering, University of Cincinnati, Cincinnati, OH 45221, USA*

*Email: bahkjg@ucmail.uc.edu

**Abstract**

Thermoelectric (TE) air cooling is a solid-state technology that has the potential to replace conventional vapor compression-based air conditioning. In this paper, we present a detailed system-level modeling for thermoelectric air conditioning system with position-dependent (graded) and constant material properties. Strategies for design optimization of the system are provided in terms of cost-performance trade-off. Realistic convection heat transfer at both sides of the system are taken into account in our modeling. Effects of convection heat transfer coefficients, air flowrate, and thermoelectric material properties are investigated with varying key parameters such as TE leg thickness, module fill factor, and input current. Both constant material properties and graded properties are considered for the TE materials, and they are compared in terms of the degree of cooling, coefficient of performance (*COP*), and power consumption. For graded materials, we employ one-dimensional finite element methods to solve the coupled electrical and thermal current equations with arbitrary profile of material properties varying with position along the TE legs. We find that graded materials can enhance the degree of cooling, but only at the expense of *COP*, compared to the case of constant property materials. With constant material properties of $ZT = 1$ and relatively low electric current, the power consumption of TE cooler can be lower than those of conventional air conditioners at an equivalent cooling capacity. Considering additional advantages such as demand-flexible operation, low noise, and high scalability, thermoelectric cooling could be a competitive technology for future air conditioning applications.

## 1. INTRODUCTION

More than 100 years has passed since the invention of the first vapor compression-based electric air conditioner in the early 1900s [1], but most contemporary houses and buildings still rely on the same technology for air conditioning. Many non-vapor compression technologies have been proposed and developed in recent years [2], which include magnetocaloric [3], thermoacoustic [4], and thermoelastic [5] technologies. However, none of them have been commercially successful thus far. Thermoelectric (TE) cooling is another promising solid-state cooling technology that can pump heat from one side of a device to the other by electric current based on the physical phenomenon called the Peltier effect.[6] It has many advantages such as no moving parts, noiseless operation, and flexible performance with control of current input. The same device can be used for power generation under heat input based on the Seebeck effect.[6]

A typical structure of a TE device is schematically shown in Fig. 1. It consists of multiple alternating p-type and n-type semiconductor TE elements, also called TE legs, which are connected electrically in series via top and bottom electrodes, and thermally in parallel, e.g. in the vertical direction in the figure. Hot air flows horizontally through a heat exchanger attached on top of the module (not shown in Fig. 1), where heat is absorbed into the TE module. The heat is then transferred to the backside of the module and dissipated through the bottom heat exchanger. During this process, the incoming air is cooled while it travels along the top of the module and the cooled air flows out of the other side. The outgoing air can pass through multiple stages of TE modules in series for sequential cooling until the desired cooling temperature is reached. The difference between final and initial air temperature is the desired cooling defined as $\Delta T_{air,target}$.

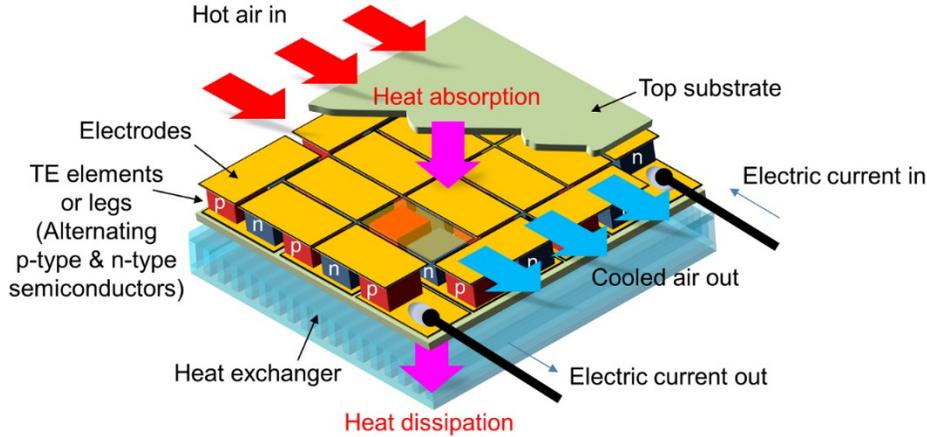

**FIG. 1.** Schematic structure of a thermoelectric air cooling module consisting of multiple TE elements with heat exchanger interfaces.

For a single-element TE cooler, the coefficient of performance (*COP*), defined as the resulting cooling power divided by the electric power input, is largely determined by the dimensionless thermoelectric figure of merit of the material used, defined as $ZT = S^2\sigma T/\kappa$, where $S$ is the Seebeck coefficient, $\sigma$ is the electrical conductivity, $\kappa$ is the thermal conductivity, and $T$ is the absolute temperature. The numerator in $ZT$, $S^2\sigma$, is called the power factor. For a multi-element module or a whole system with heat exchangers like the one shown in Fig. 1, however, the overall *COP* highly depends on several module design parameters such as element thickness, fill factor (the fractional coverage by TE elements), as well as the convective heat transfer characteristics at both sides. Therefore, careful design optimization with these parameters is essential to achieving optimal cooling. In addition, there is a trade-off between cooling and power consumption for a TE cooler as similarly shown in TE power generators between the power generation efficiency and cost.[7] Thus a thorough analysis about the trade-off between performance and cost is important for a cost-effective design optimization. A realistic simulation of a TE cooler that can take into account the impacts of external convection heat transfer, module design parameters, and material properties on various performance measures is therefore very useful at the design stage.

While modeling a TE device, it is also important to note whether to model the properties as being constant or changing, for example, with position or temperature. For a large temperature difference across TE elements, temperature-dependent material properties must be taken into account. In air conditioning applications, however, there is a relatively small temperature difference close to ~ 10 K, such that a constant property model (CPM) may work well. In CPM, the Peltier effect is localized at the interfaces of TE materials, e.g. interfaces with the electrodes as no Thomson effect occurs inside the leg, so that TE module simulation is relatively straightforward. In CPM, the maximum cooling for a single-element TE cooler is known to be limited by the material figure of merit, such that $(\Delta T)_{max} = ZT_c^2/2$, where $T_c$ is the cooled side temperature [8].

There have been several recent efforts to beat this maximum cooling limit by utilizing segmented or graded materials. In 2006, Bian et al. [8] theoretically showed that the maximum cooling can be improved by segmenting the TE leg to have an increasing Seebeck coefficient along the length of the TE leg from cold to hot side while keeping the power factor and *ZT* the same. At the expense of *COP*, the maximum cooling is improved. Since the ZT is unchanged, the cooling enhancement is attributed to the redistribution of Peltier and Joule heating inside the TE leg. They later found that at optimal grading, the maximum cooling can be improved by 27 % in $Bi_2Te_3$ material [9]. Zhou et al.[10] showed numerically that further enhancement of cooling is possible by using pulsed operation in a graded cooler. Snyder et al.[11] used the concept of self-compatibility in designing optimal graded materials for a single-element TE cooler, which showed approximately twice the maximum cooling compared to the CPM case with an equivalent *ZT*. Later, the two different graded coolers proposed by Bian et al. [8] and Snyder et al. [11] were systematically compared for various materials and operating conditions by Seifert et al. [12]. Jin et al. [13] showed an increase in energy conversion efficiency of a functionally graded TE module using an exponential variation of *ZT*. In their case, *ZT* was higher at the hot end of the module. A comprehensive study of cooling power ($Q_c$) in a graded model is done by Seifert et al. [14]. They showed an increase in cooling power density compared to a CPM for various design of graded materials. The concept of graded materials is equally applicable to a TE generator [15]. Recently, Thiébaut et al. [16] obtained an analytical solution for a graded cooler that maximizes the degree of cooling under the assumption of a small cooling, i.e. $\Delta T \ll T_H$, where $T_H$ is the hot side or ambient temperature. The solution is given as the form of a differential equation for the relationship between the resistivity and the Seebeck coefficient, and then the closed form solutions are provided for the cases of typical nondegenerate semiconductors.

All these previous theoretical studies on graded coolers, however, mainly focused on maximizing the degree of cooling for a single-element device with unrealistic external heat transfer conditions such as an adiabatic condition at the cooled side with a perfect heat sink at the hot side. The impacts of the graded scheme on *COP* have been relatively unexplored, and the design optimization with multi-element modules and realistic external heat transfer coefficients has not been studied significantly either. In a large-scale air conditioning application, a number of multi-element TE modules must be used to cover a large heat exchange area. Also, keeping *COP* high is essential in a cost-effective cooling operation because the power consumption, and thus the operational cost, is determined by *COP* once the desired cooling power is reached. In this work, we propose to use a module design that keeps the *COP* sufficiently high at the expense of the degree of cooling per module as there is a trade-off between the two performance measures. Yet, the target degree of cooling can be met if several cooling modules are used in series for sequential cooling. Although the cost will increase with increasing number of modules used, we find that the *COP* can increase disproportionately, such that the overall cost can be maintained or even lowered for a target degree of cooling.

Recently, a lot of thermoelectric A/Cs or air cooling designs have been experimentally and numerically analyzed. Aklilu et al. reviewed the integration of a TE AC system in a building involving radiant panels, air and water ducts [17]. An inverse relation between COP and the degree of cooling was shown, which was also observed experimentally. They had a COP on the order of 2.5 for a hot to cold side temperature difference of about 15 K. Similarly, focus on integration and on improving COP was done by Xiaoli et al. [18]. Their techniques involved ventilated façade for heating, cooling and heat recovery etc. Anthony et al. performed theoretical and experimental analysis on a small powered TE device [19]. They had a small COP of 0.465 at a cooling temperature difference about 4.5 K, and the device had a low power compared to a conventional A/C. Hamed et al. introduced new techniques to improve COP of a TE cooling device using jointed modules and improved heat sinks [20]. Recent developments in models and materials for TE power generation and cooling are summarized in a review by Pourkiaei et al. [21]. Numerical simulation of a tubular TE module in air cooling device was investigated by Tian et al. [22]. Their COP was on the order of 1, depending on various parameters. Irshad et al. reported energy saving in a TE cooling device compared to a conventional AC despite having lower COP of 0.679 and getting a cooling of 3-5 K [23]. Two-way heating and cooling TE devices were experimentally tested in realistic conditions by Maria et al. [24] and by Yilmazoglu [25]. Maria et al. [24] achieved a cooling COP up to about 1.7 for a cooling of 15 K while Yilmazoglu [25] had a COP less than 1 at a cooling of 4 K. An air-water based TE cooling system was experimentally studied by Hamed et al. [26]. The COP in their case reached up to 3 and the cooling is greater than 30 K. An air-air system was designed and tested by Attar et al. [27], where the COP reached up to 1.27 and the cooling of air was 15-20 K. Numerical simulation of a small TE cooler using three-dimensional FEM was conducted by Gong et al. [28]. Their COP reached up to 2.5 depending on various parameters at a hot-to-cold side temperature difference of 10 K. In most of these works, the COP was low (less than 1). Hence, in the present work our aim is to carefully select a set of parameters to get a COP as high as 4, as explained in the results. Also, our target cooling is kept constant at 10 or 13 K.

In this work, we first provide the detailed modeling scheme for multi-element TE cooler modules with realistic external heat transfer coefficients for large-scale air conditioning. Then we discuss the trade-off relation between the degree of cooling and *COP* with the calculation results. Finally, we provide strategies for designing cost-effective TE air conditioning systems that can potentially beat the conventional air conditioning technology in terms of *COP* and cost.

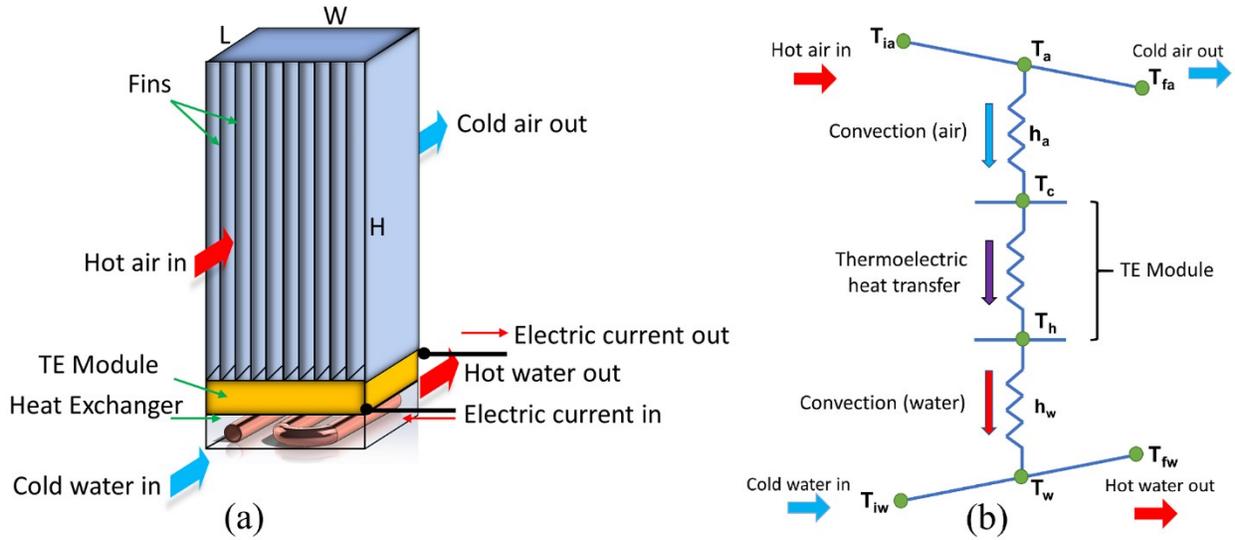

**FIG. 2.** (a) Schematic diagram of a TE module of 5 x 5 cm$^2$ area with heat exchangers. Air flows through the shown channels of the plate-fin heat exchanger with height H= 10 cm, length = 5 cm, and width = 5 cm. Due to fins, the width is divided into 10 sections each of about 5 mm width. On the other side of TE device, water flows in serpentine shaped pipes for heat removal. When current is given to the device, heat is extracted from air and rejected to water. There can be multiple modules in both directions: along and transverse to the flow of air. (b) Air, water and TE leg hot and cold temperatures at different points in a TE leg for a CPM. The temperatures $T_a$, $T_c$, $T_h$, and $T_w$ are obtained by solving the heat balance equations (Eq. 1,2,13 and 14) while initial air and water temperatures are assumed.

## 2. SYSTEM AND THEORETICAL MODEL

### 2.1. Modeling of external convection heat transfer

Fig. 2(a) shows our module design with heat exchangers. This device is applied as an air cooler, where inlet air at room temperature is expected to be cooled by the device for space cooling. On the cooled side, a plate fin-type heat exchanger is used and the air passes through the channels between the fins for efficient heat exchange. On the hot side, water is used as a heat absorber because of its good specific heat. To connect the air and water ducts to the TE module, we assume a very high thermally conducting material with no additional interface thermal resistance to reduce heat loss.

We also assume that the temperature on the outer surface of the water pipe is varying only with length, so the material connecting the pipe and TE legs can be taken as having negligible thermal resistances. The number of modules along the flow of air is calculated based on the desired air temperature drop (10 and 13 K). For one module, the total heat transfer is the sum of heat transfer in the individual legs.

The dimensions of the air channel and water duct in one module are also shown in Fig. 2(a). Air and water enter their respective ducts at an inlet temperature $T_i$ (300 K) and exit at temperature $T_f$. Since the temperature change of the fluids are relatively small, we can assume that the top and bottom side surfaces of the TE module are maintained at uniform temperatures, $T_c$ and $T_h$, at the cold and hot sides, respectively. Fig. 2(b) shows the temperatures involved in the heat transfer equations. The heat transfer $Q_c$ occurs from air to module, and $Q_h$ from module to water as

$$h_a A_t (T_a - T_c) = Q_c = \dot{m}_a C_{pa}(T_{ia} - T_{fa}) \quad (1)$$

$$h_w A_t (T_w - T_h) = Q_h = \dot{m}_w C_{pw}(T_{iw} - T_{fw}) \quad (2)$$

where, $h$ is the heat transfer coefficient, $A_t$ is the area of the module, $C_p$ and $\dot{m}$ are the specific heat capacity and the mass flow rate respectively. The subscripts $a$ and $w$ are used for properties related to air and water respectively. $T_a$ and $T_w$ are the average temperatures of air and water respectively calculated by Eq. (3) and Eq. (4). The air temperature decreases, and water temperature increases along the direction of flow. For simplicity, we can assume a linear variation of the temperature, so that the average heat transfer across the module area is given by the left-hand side of Eq. (1) and Eq. (2), which, by energy conservation, is equal to the internal heat loss of air given by the far right-hand side of Eq. (1) and Eq. (2).

$$T_a = (T_{ia} + T_{fa})/2 \quad (3)$$

$$T_w = (T_{iw} + T_{fw})/2 \quad (4)$$

where $T_i$ and $T_f$ are initial and final temperatures respectively for one module, and the subscripts $a$ and $w$ denote air and water, respectively.

An important part of the mathematical formulation is to estimate the heat transfer coefficients for both air and water flow. As shown in Fig. 2, we used 10 channels between fins for air flow. For each channel, the hydraulic diameter is given by

$$D_h = \frac{4A}{P} \quad (5)$$

where $A$ and $P$ are the cross-sectional area and the perimeter of such a channel, respectively. For a rectangular cross section as is the case for the channel, the Nusselt number Nu for a fully developed laminar flow depends on the $b/a$ ratio denoted as α [29], where $b$ and $a$ are, respectively, the shorter and longer edges of the rectangle. This ratio is 1/20 for the present channel design (Fig. 2(a)), for which Nu is 6.6 calculated by Eq. (6) provided by Shah and London [29]. This value is for constant wall temperature case, but in our model, there is a combination of two cases: 3 walls have constant temperature and one has constant heat flux. Since this is complex to determine and the fact that Nu is usually higher for a constant flux case, we conservatively take the Nusselt number as 6.6. Once the Nusselt number is known, the heat transfer coefficient can be calculated from Eq. (7).

$$Nu_a = 7.541(1 - 2.61\alpha + 4.97\alpha^2 - 5.119\alpha^3 + 2.702\alpha^4 - 0.548\alpha^5) \tag{6}$$

$$h_a' = \frac{Nu_a k_a}{D_h} \tag{7}$$

where $k_a$ is the thermal conductivity of air. Note that $h_a'$ is the average heat transfer coefficient per the channel surface area. The effective heat transfer coefficient per the module area can be calculated using Eq. (8) by accounting for the surface area enhancement by the fins:

$$h_a = \frac{A_s h_a'}{A_t} \tag{8}$$

$$A_s = 2n_s H L \tag{9}$$

where $A_s$ is the surface area of the entire heat exchanger in contact with air, $A_t$ is the area of one TE module, $H$, and $L$ are respectively the height and length of the channel. $n_s$ is the number of channels, 10 in this case. The factor of 2 in Eq. (9) accounts for two surfaces per channel.

On the water side, we take a fully developed turbulent flow as the Reynold's number, $Re = \frac{4\dot{m}_w}{\pi \mu D}$ is much greater than the transitional Re ~ 2,300 (Table 2). Here, µ is the water viscosity. We use the Dittus-Boelter equation (Eq. (10)) for the Nusselt number for a turbulent flow [30], which depends on the Reynold's number and Prandtl number Pr as shown:

$$Nu_w = 0.023 Re^{0.8} Pr^{0.3} \tag{10}$$

The heat transfer coefficient can be calculated in the same way as air, by using the thermal conductivity $k_w$ and pipe diameter $D$, such that

$$h_w' = \frac{Nu_w k_w}{D} \tag{11}$$

The effective heat transfer coefficient for water is

$$h_w = \frac{A_{sw} h_w'}{A_t} \tag{12}$$

where $A_{sw}$ is the surface area of water pipe given by

$$A_{sw} = \pi D L' \tag{13}$$

$L'$ is the length of water pipe in contact with the TE module, and it can be increased by using a serpentine pipe as shown in Fig. 2.

The thermophysical properties of air and water are taken at 20 and 25 °C, respectively [31][32], based on the average temperature they are at during their contact with the TE module. The values are given in Table 1. Based on the mass flow rate and duct geometry along with the thermophysical properties, heat transfer coefficients for both air and water are calculated using the above equations. Reynold's number and effective heat transfer coefficients as calculated by Eqs. (8) and (12) are shown in Table 2. A better heat transfer coefficient with a lower mass flow rate may result in a larger temperature drop for air, which is desirable. The air mass flow rate shown is for one module, and it increase with the number of modules used in parallel.

**Table 1.** Thermophysical properties of air and water. These properties vary with temperature, so they are taken at average temperatures of 20 and 25 °C for air and water, respectively, which these fluids are at during their flow. All properties are at atmospheric pressure.

| Property | Air (20 °C) [31] | Water (25 °C) [32] |
|---|---|---|
| Specific Heat $C_p$ (J/kgK) | 1007 | 4183 |
| Thermal Conductivity $k$ (W/mK) | 0.025 | 0.59 |
| Dynamic Viscosity $\mu$ (Ns/m²) | 1.82 X 10$^{-5}$ | 8.9 X 10$^{-4}$ |
| Density $d$ (kg/m³) | 1.20 | 997.1 |
| Prandtl Number $Pr$ | 0.73 | 6.26 |

**Table 2.** Flow parameters. The effective heat transfer coefficient and Reynold's number are calculated based on fluid properties, geometry and flow conditions. Former is a very important parameter to be considered while considering heat transfer between TE device and fluids. Both are internal flow for which transition Reynold's number is roughly 2300. Therefore, air flow is laminar and water flow is turbulent. Air mass flow rate is shown for one module.

| Variable | Cold side (Air) | Hot side (Water) |
|---|---|---|
| Heat transfer coefficient (W/m²K) per module area | 590.98 | 12577.5 |
| Reynold's number | 1465.1 | 14298 |
| Mass flow rate (kg/s) | 0.002 | 0.05 |
| Hydraulic/Duct Diameter (cm) | 0.95 | 0.5 |

## 2.2. Thermoelectric cooler with constant material properties

We first model a thermoelectric cooler module with the constant property model (CPM). For one pair of n-type and p-type legs, the cooling power $Q_c$ and $Q_h$ are given by

$$Q_c = ST_c I - \frac{1}{2} I^2 R - K(T_h - T_c) \tag{14}$$

$$Q_h = ST_h I + \frac{1}{2} I^2 R - K(T_h - T_c) \tag{15}$$

Here, the first term is the Peltier cooling, and the second and the third terms indicate, respectively, Joule heating and thermal conduction. $S$, $R$, and $K$ are, respectively, the Seebeck coefficient, the electrical resistance, and the thermal conductance of one pair of TE legs given by

$$S = S_p - S_n \tag{16}$$

$$R = \frac{L_l}{\sigma_p A_l} + \frac{L_l}{\sigma_n A_l} + R_c \tag{17}$$

$$K = \frac{k_p A_l}{L_l} + \frac{k_n A_l}{L_l} \tag{18}$$

where, $L_l$ is the TE leg thickness, $A_l$ is the TE leg area and subscripts $p$ and $n$ indicates the p-type and n-type material properties, respectively. Note that $S_p > 0$, and $S_n < 0$, so that their magnitudes are added in the total Seebeck coefficient, which indicates that both types constructively contribute to the Peltier cooling. $R_c$ is the additional resistance from electrodes and contacts. We use a constant $R_c = 0.01$ Ω per a pair of TE legs [33]. $I$ is the input current applied in the device and most of the optimization is done to find an appropriate current.

For the CPM case, we conduct simulations with a material having $ZT=1$ at 300 K, for which we selected the individual properties similar to those of $Bi_2Te_3$ as summarized in Table 3.

**Table 3.** Thermoelectric properties of the material used having a ZT value of 1 at 300 K. We assume the same magnitude of properties for both p-type and n-type for convenience. To estimate performance for a different ZT later, the Seebeck coefficient is changed as shown in Table 5, while keeping the other properties the same.

| | |
|---|---|
| Magnitude of Seebeck coefficient $S$ (µV K$^{-1}$) | 258.2 |
| Electrical conductivity $\sigma$ (Ω$^{-1}$ m$^{-1}$) | $5 \times 10^4$ |
| Thermal conductivity $k$ (W m$^{-1}$ K$^{-1}$) | 1.0 |
| ZT | 1.0 |

Fill factor is an important parameter for a TE module design. It is defined as the fractional coverage of TE elements for the entire module area, such that

$$F = \frac{2NA_l}{A_t} \quad (19)$$

where $N$ is the number TE leg pairs. We design a TE module in the way that the fill factor is selected, and then the number of TE leg pairs, $N$, is determined by the fill factor with fixed leg cross-sectional area and module area. These parameters and TE leg thickness alter the way a TE device performs. Hence geometry optimization is done to find a suitable thickness and a fill factor.

Table 4 shows additional TE module parameters chosen for the module simulation. The parasitic heat conduction through gap fillers is often neglected in TE module simulation, but it can be significant when the fill factor is small. In our model, we assume an air gap, and the gap filler thermal conductivity $k_g$ of air is incorporated in the total pair thermal conductance $K$ as

$$K = \frac{k_p A_l}{L_l} + \frac{k_n A_l}{L_l} + \frac{k_g 2 A_l (1-F)}{L_l F} \quad (20)$$

where the factor $2A_l(1-F)/F$ is the gap area per a pair of legs in the module. Thermal contact resistance is difficult to determine because of the complexity of the heat transfer. It has been addressed in the past [34], however, in the present work, thermal contact resistance is neglected.

**Table 4.** TE module parameters. Gap filler thermal conductivity can be significant for low fill factors, so it should be considered. This value is taken assuming air filling the gap between TE legs. While keeping the values given in this table constant, Leg thickness and Fill factors are used as design parameters to optimize the cooling and efficiency (*COP*) of the device.

| Gap Filler thermal conductivity $k_g$ (W m$^{-1}$ K$^{-1}$) | 0.026 |
|---|---|
| Module area $A_t$ (cm$^2$) | 25 |
| TE leg area $A_l$ (mm$^2$) | 4 |

For CPM, the thermoelectric properties are constant along the length of TE leg. So, we have 4 equations (Eqs. 1,2,14,15) and 4 unknown temperatures; $T_a$, $T_w$ (Average air and water temperature), $T_c$ and $T_h$. These are solved for different currents using the Gauss Elimination method for simultaneous linear equations [35]. Once we have the temperature profile, we can find the *COP*, and the cooling power given by Eq. (14) or more simply the temperature drop of air, also called the degree of cooling, $\Delta T_{air}$, such that

$$COP = \frac{Q_c}{I(S(T_h - T_c) + IR)N} \tag{21}$$

$$\Delta T_{air} = T_{ia} - T_{fa} \tag{22}$$

The denominator in the *COP* equation, Eq. (21), is the total power input for the entire module so the power input to one pair is multiplied by the number of pairs *N*. For one pair, the external voltage supply must overcome the Seebeck voltage, $S(T_h - T_c)$, created against the external voltage due to the temperature difference along the legs, which is added to the ohmic voltage, *IR*.

## 2.3. Thermoelectric cooler with graded materials

For a graded material, we adapt the analytical solution given by Thiébaut et al. [4] for maximum degree of cooling. The Seebeck coefficient *S* and the resistivity *ρ* vary with position in a TE leg as given by

$$S(x) = S_0 + \frac{k_B}{e} \ln\left(\frac{1}{1 - \frac{x}{L_l}(1-c)}\right) \tag{23}$$

$$\rho(x) = \rho_0 \frac{1}{1 - \frac{x}{L_l}(1-c)} \tag{24}$$

where the position *x* is along the length of a TE leg, with $x=0$ at the cold side, and $x=L_l$ at the hot side, $k_B$ is the Boltzmann constant and *e* is the electron charge. In this model, the two properties vary with *x* via position-dependent doping level in the semiconductor. The parameter *c* ($0 < c < 1$) is used to describe how much the doping varies with position and is defined as the ratio of doping concentration between the hot and cold sides. Between the two ends, the doping concentration is assumed to change linearly with position. For instance, $c=1$ for the same doping along the leg, which is the CPM case. The smaller the *c* is, the larger the doping varies between the two ends. In this work, *c* is taken as 0.1 to provide reasonable doping change along the leg for a significant grading effect. Lowering the *c* value further would not produce much significant improvement [16].

To take into account the position-dependent material properties, we employ a 1D finite element method. The temperature profile along a TE leg is solved for a number of nodes *n* in the leg as shown in Fig. 3(a). The heat balance equation for an internal node is formulated based on the thermal network model shown in Fig. 3(b), and given for the *i*-th node by

$$(S_{i+1} - S_i)T_i I = (T_{i-1} - T_i)K_i - (T_i - T_{i+1})K_{i+1} + \frac{1}{2}I^2((R_{i+1} + R_i)) \qquad (25)$$

where the left-hand side is the Peltier heat denoted by $Q_P$ in Fig. 3(b), which is balanced by the Joule heating and the conduction terms ($Q_J$ and $Q_K$).

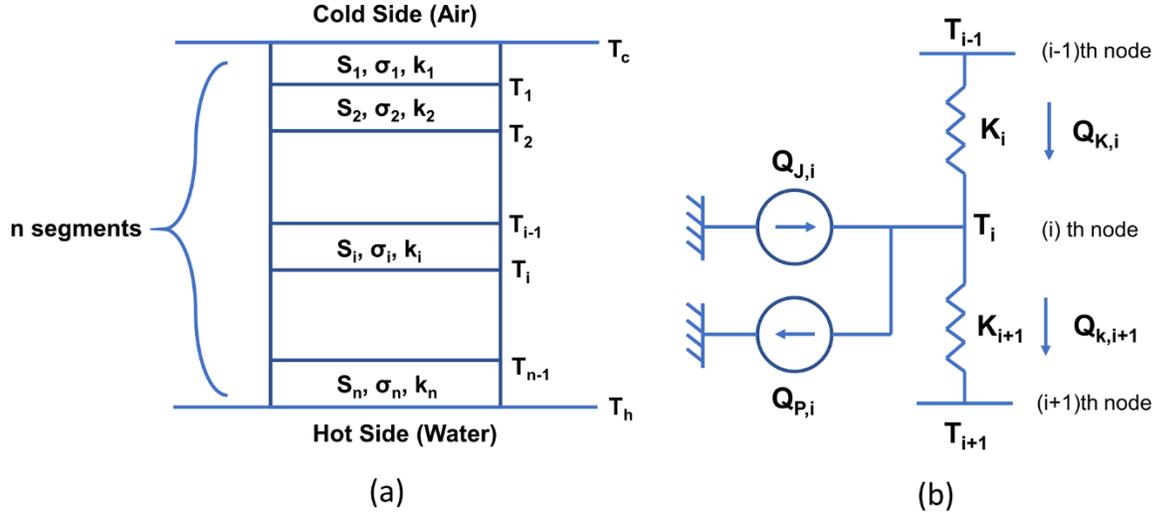

**FIG. 3.** (a) Nodes on one TE leg for modeling a graded material. The TE properties, Seebeck coefficient, and electrical resistivity vary along the length, hence temperature is calculated at different nodes shown here. (b) Components of heat transfer on the $i$-th node. $Q_J$, $Q_P$ and $Q_K$ are Joule, Peltier and Conductive terms respectively. Peltier cooling is compensated by Joule heating and heat conduction. To obtain cooling, an optimum current should be considered to minimize Joule and conduction heating terms.

Now, the resistance and thermal conductance are divided into $n$ nodes in series, so at the $i$-th node:

$$R_i = \frac{L_l}{\sigma_i n A_l} \qquad (26)$$

$$K_i = \frac{n k A_l}{L_l} \qquad (27)$$

Then, the total resistance in a leg is the sum of all the nodal resistances and the electrical contact resistance such that

$$R = \sum_{i=1}^{n} R_i + Rc \qquad (28)$$

Thermal conductance includes the gap filler thermal conductivity calculated in a similar manner as in CPM. Other parameters are the same as in CPM. Since geometry optimization is done for CPM only, the obtained leg thickness and fill factors are used in both CPM and graded, while making a comparison for different *ZTs*. The *COP* is obtained in a graded device by

$$COP = \frac{Q_c}{I(V_s + IR)N} \qquad (29)$$

where the denominator is the power input for a module, $W_m$ and $V_s$ is the total Seebeck voltage created in the leg given by

$$V_s = \sum_{i=1}^{n} S_i(T_i - T_{i-1}) \qquad (30)$$

For either CPM or graded materials, we can find the required number of modules $N_m$ along the flow of air by taking into account the degree of air cooling produced by one module, $\Delta T_{air}$, and the total desired cooling $\Delta T_{air,target} = 10$ or 13 K in our case such that

$$N_m = \frac{\Delta T_{air,target}}{\Delta T_{air}} \quad (31)$$

Total electrical power input (or power consumption) is then calculated for the entire system as

$$W_{tot} = N_{pm} N_m W_m \quad (32)$$

where $N_{pm}$ is the number of parallel modules, which is estimated based on total air mass flow rate requirement. Finally, the electricity cost per hour is calculated by multiplying the total power consumption by the average electricity price in the unit of $/kWh. Although, there will be an additional power requirement for pumping the fluids, the total cost can still be lower than that of conventional A/Cs. This is explained more in section 4.4.

## 3. SIMULATION PROCEDURE

The mathematical model from Section 2 is coded in MATLAB for both CPM and graded (GPM) materials. The first step is to optimize one CPM module based on the air temperature drop and *COP*, with different leg thickness and fill factors, and varying electrical current. After the optimization, we select the desired thickness and fill factor, which are eventually used in further simulations. Heat balance equations produce a system of simultaneous linear equations which are solved using the Gauss Elimination method [35]. This provides an exact solution for the temperature profile. In a CPM, there are only 4 equations because temperature is calculated only at the ends of TE legs (Eqs. (1) and (14)). Also, in the CPM calculations, Seebeck coefficient *S* is varied to alter the ZT from 0.8 to 1.4 while keeping thermal conductivity and electrical resistivity the same (Table 5).

Fig. 4 (a), (b) and (c) show the variation of Seebeck coefficient, resistivity and *ZT*, respectively, along the length of the TE leg for $ZT_{max} = 1$ in a graded segment. The thermal conductivity is assumed to be constant as the temperature range is narrow (290-310 K) and it would not vary significantly. Cooling and *COP* calculations are done for various reasonable values of $ZT_{max}$ that are in accordance with commercially available TE materials (0.8, 1.0, 1.2 and 1.4). To alter the *ZT*, we vary the Seebeck coefficient $S_0$ at the cold side of the segment. Thus, *S* varies with both position and ZT (Fig. 4(a) and Table 5). Although, in reality both Seebeck coefficient and resistivity are inter-related and vary with each other, here we assume resistivity varies only with length and not with ZT while keeping thermal conductivity constant throughout. In a GPM, due to the varying TE properties in the graded segment, temperature is solved at each node along the length of TE leg. This also causes *ZT* to be different at each node. To make a comparison with CPM, the maximum *ZT* in the GPM leg is compared with the constant the ZT value of the CPM.

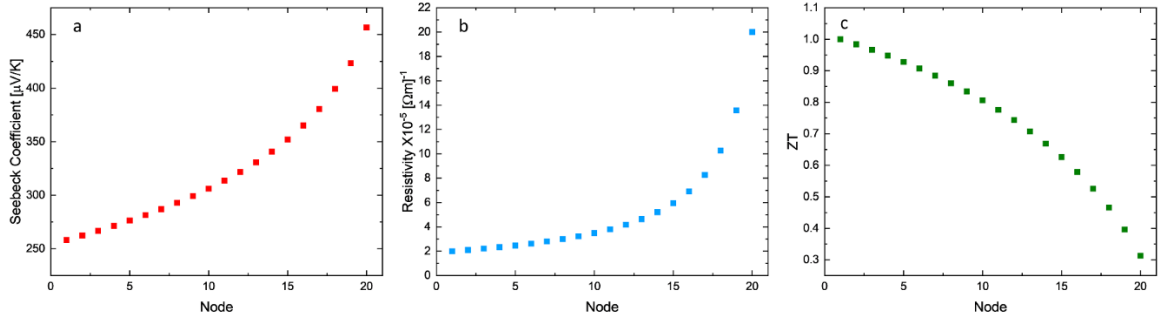

**FIG. 4** Variation of TE properties along the length of a TE leg. The values are taken from Eqs. (23) and (24) [4]. (a) Seebeck coefficient has a value of 258.2 μV/K at the first node ($S_0$), which is at the cold side of the device, and vary along the leg following Eq. (23). (b) Resistivity is $2 \times 10^{-5}$ $\Omega^{-1}$ m$^{-1}$ at the same node ($\rho_0$) and vary along the leg thickness following Eq. (24). Variation for Seebeck will be different for a different value of ZT because of different $S_0$ (Also shown in Table 5). (c) ZT with the maximum ZT of 1 at the first node. The thermal conductivity was assumed to be 1 W m$^{-1}$ K$^{-1}$.

**Table 5.** Seebeck coefficient for different values of ZTs. For a CPM, this value will be constant for a TE leg. However, for a graded segment, these values are only at the cold side of the leg ($S_0$). The variation can be seen in Fig. 4(a) and (c).

| Seebeck coefficient (μV/K) $S$ (CPM) or $S_0$ at the cold-side boundary in a graded segment | ZT at the cold-side boundary in a graded segment (constant for CPM) |
|---|---|
| 230.94 | 0.8 |
| 258.2 | 1.0 |
| 282.84 | 1.2 |
| 305.5 | 1.4 |

Next consideration is the electrical power consumption. We compare the power consumption and COP with typical A/Cs. By a little extra expense on the initial (material) cost, we can reduce the electricity cost by using a TE cooler. The number of modules along the flow of air is calculated by dividing the objective air temperature drop (10 or 13 K) by the temperature change calculated for one module. Total number of modules can then be found by estimating the number of parallel modules required for a target air flow rate. Effect of air flow rate is analyzed by varying either the mass flow rate of air through on series of modules or the total mass flow rate for whole device as explained in sections 4.3. We aim to provide air cooling at a lower electricity cost than conventional air conditioners.

## 4. RESULTS AND DISCUSSION
### 4.1. Geometry Optimization in CPM at a ZT of 1

In the first part, we optimize the module geometry with TE leg thickness and module fill factor. This part is only for a CPM, and we take an example of $ZT = 1$. Fig. 5(a) shows $\Delta T_{air}$ (air temperature drop across one module) variation with current at different leg thicknesses of 0.5, 1, 2, and 3 mm, while keeping the fill factor constant at $F = 0.1$. At lower currents (less than 2 A), the effect of leg thickness is not significant (except for 0.5 mm), which is mainly due to the trade-off between the electrical and thermal resistances.

Since the Joule heating term varies as $I^2$, the effect of TE leg thickness is more profound at higher currents. Due to the increased electrical resistance and current, which together increase the detrimental Joule heating inside the TE leg, cooling is decreased for thicker legs at the high current region. There is an optimum current that maximizes the degree of cooling, which is due to the interplay between Peltier cooling and Joule heating, both of which depend on current. This optimum current decreases with increasing leg thickness because of increasing electrical resistance, which needs a lower value of current to keep the Joule heating sufficiently low.

Fig. 5(b) shows the *COP* variation with the same setting. *COP* monotonically decreases with increasing leg thickness over the entire current range. *COP* is the ratio of cooling power to electrical power input, and the power input directly increases with resistance and current. Higher leg thickness increases resistance and thus the power input. Hence, the effect on *COP* is a strictly decreasing function of leg thickness and of current. To select a particular value of leg thickness for design, both material saving (smaller leg thickness) and ease of manufacturing (not too small of leg thickness) can be considered. Therefore, 2 mm can be a suitable value. Further design optimization is done for the leg thickness of 2 mm. Note that this leg thickness is conservatively chosen, as a smaller thickness, e.g. 1 mm, can achieve even a higher *COP* with a similar degree of cooling but is significantly more difficult to manufacture.

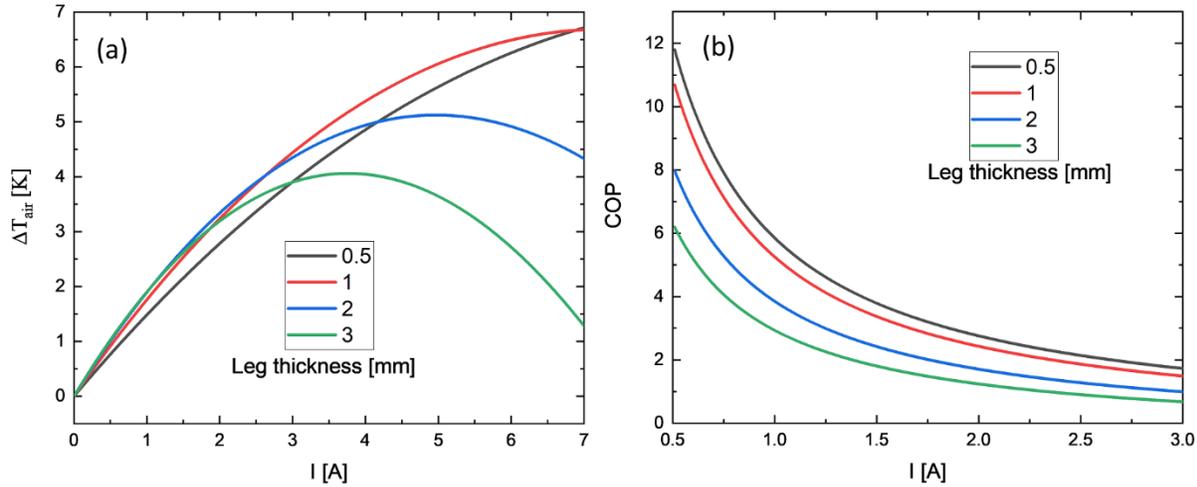

**FIG. 5.** Variation of (a) the degree of air cooling ($\Delta T_{air}$) and (b) the coefficient of performance (*COP*) of a $5 \times 5$ cm$^2$ TE module with current at different values of leg thickness. The fill factor is kept constant at $F=0.1$. Selection of a thickness as a design parameter may depend on whether cooling or efficiency is being considered, as is clear from this figure, that larger thickness gives a smaller *COP* but offers similar $\Delta T_{air}$ to thinner legs at lower currents.

Fig. 6 (a), (b) and (c) show the $\Delta T_{air}$, *COP* and power input, respectively, for one $5 \times 5$ cm$^2$ module with varying current at different fill factors (0.1, 0.2, 0.4, 0.6, 0.8) at a constant leg thickness of 2 mm. It is noticed that the degree of cooling increases for increasing fill factor over the entire current range. This is due to the increasing number of TE leg with increasing fill factor with a constant leg area, causing more total Peltier cooling. But the increasing fill factor increases electrical resistance at the same time, which increases the power input, and thus reduces *COP* significantly. As a result, *COP* monotonically decreases with increasing fill factor for this particular leg thickness as shown in Fig. 5(b). The optimum value of current for maximum degree of cooling is between 4 and 5 A, but we are only concerned with smaller currents below 2 A for cost-effective operation: at those high currents, the *COP* is too low, and the power input is too high. To reduce the electrical power consumption and hence electricity cost, we select a lower value of *F* as 0.1 and smaller currents (< 2 A). At these conditions, the degree of cooling is around 3 K or less, so we need just 3 modules in a series or a little more to achieve total cooling greater than 10 K. (Additionally, there will be a number of such series in parallel based on the required mass flow rate which is discussed in Section 4.3). So, we select a leg thickness as 2 mm and a fill factor of 0.1. Higher fill factors would have increased cooling but due to the decreasing *COP*, power input is higher and hence power consumption increases significantly as shown in Fig. 6(c). Smaller fill factors and thinner legs could cause substantial heat and current spreading and contraction at the interfaces with substrates as well as increased parasitic heat loss through the fillers, which would require more rigorous three-dimensional analysis.

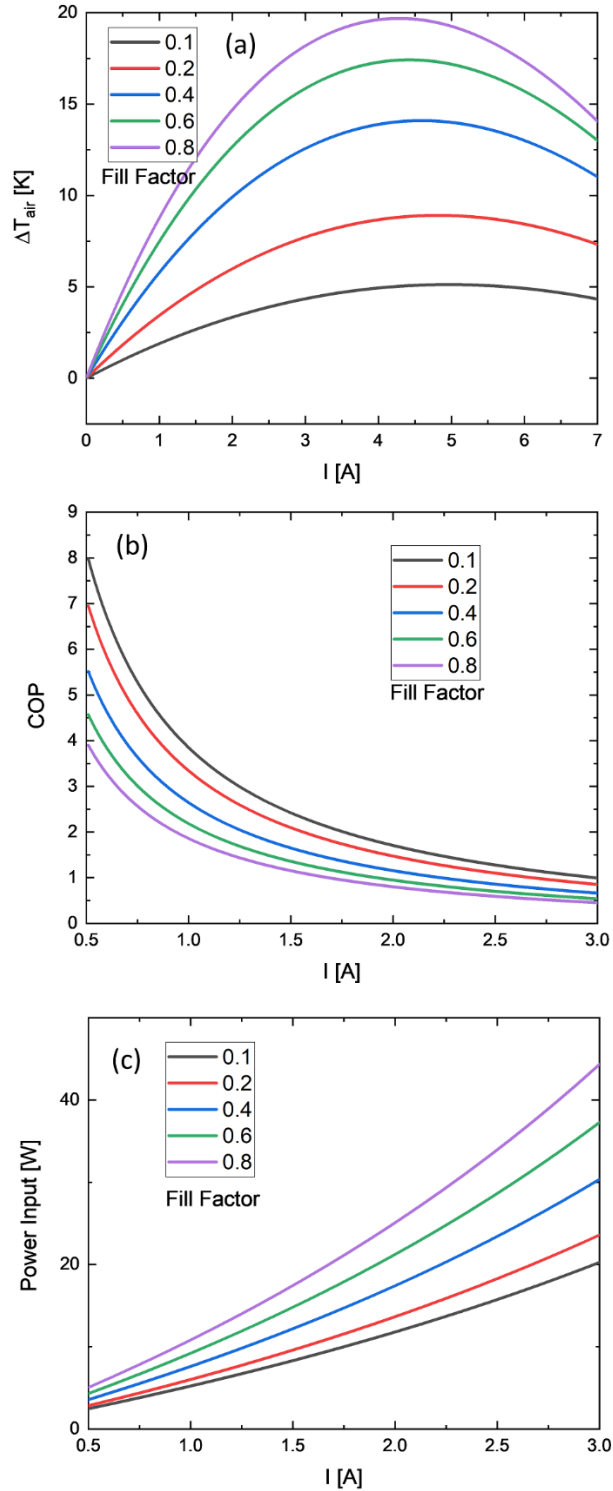

**FIG. 6.** Variation of (a) the degree of air cooling, (b) *COP* and (c) electrical power consumption or power input for a $5 \times 5$ cm$^2$ module with current for different fill factors $F$=0.1~0.8 at a leg thickness of 2 mm. Power is calculated for one series of modules. As is clearly visible, increasing $F$ produces more cooling at the expense of *COP*, and hence more power consumption.

## 4.2. Graded TE legs

Once the geometry is fixed, the performance of CPM can be compared with graded TE legs. Eqs. (23) and (24) show the equations governing the Seebeck coefficient and resistivity, respectively, as a function of distance along a TE leg for the graded material. The boundary conditions at the cold-side boundary, $S_0$ and $\rho_0$, are chosen to keep the maximum $ZT$ varying as 0.8, 1.0, 1.2 and 1.4 in order to make a comparison with CPM. $ZT$ variation along the length of a TE leg in the case of $ZT_{max} = 1.0$ is shown in Fig. 4(c).

Fig. 7(a) shows the degree of cooling $\Delta T_{air}$, as a function of current at different $ZTs$ for CPM and graded models. Fill factor and TE leg thickness are taken as 0.1 and 2 mm, respectively, from the geometry optimization part. First thing to notice is that $\Delta T_{air}$ increases with $ZT$ for both CPM and GPM, which is expected due to better thermoelectric properties. Due to higher Seebeck coefficient, Peltier cooling increases especially since resistivity and thermal conductivity are kept the same for different $ZTs$. Furthermore, the graded model enhances cooling at all current values, which is attributed to the redistribution of Peltier cooling inside the TE leg with varying Seebeck coefficient with position [8]. The increasing Seebeck coefficient with position induces Peltier cooling rather than Peltier heating inside the TE leg, which cancels the internal Joule heating to make the temperature profile linear, and thus achieve a larger cooling that that of CPM.[16]

Fig. 7(b) shows the *COP* as a function of current for the same set of different ZT values in CPM and GPM. As shown in the figure, in all cases, *COP* increases rapidly with decreasing current. Hence, a low current operation is justified for efficient TE cooling. The variation of *COP* with $ZT$ at a fixed current is relatively small compared to that of the degree of cooling. For higher $ZT$, *COP* is higher as expected due to the higher cooling power. But since power input also increases (Fig. 7(d)), the effect is not that significant. Comparing CPM with GPM, all the CPM cases show much higher *COP* than the GPM regardless of the ZT value of this range at all currents. For instance, even the *COP* for CPM at $ZT = 0.8$ is higher than that of GPM with $ZT = 1.4$ at lower currents. *COP* of the GPM is almost 50 % lower than that of CPM especially at low currents. Although, the reason that we are getting lower *COP* for graded material is that the $ZT$ for CPM is constant along the length of TE leg, while in graded it decreases sharply as seen from the Fig. 4(c) because that trend has been chosen for maximizing the degree of cooling.[16] This is visible in Fig. 7(a) where we can see an improvement of about 2 K in cooling, by grading the TE leg. The average $ZT$ is lower in graded case and could be the reason *COP* is lower for graded material as *COP* is directly related to the $ZT$ of the material.

Number of modules required in a series are calculated directly from the degree of cooling (Eq. (31)) to get a target cooling of 10 K for the device as a whole. Hence, the required number of modules decreases with current since cooling is increased (Fig. 7(c)). Any variation in this number due to the ZT or model would have the same reasoning as given for the degree of cooling.

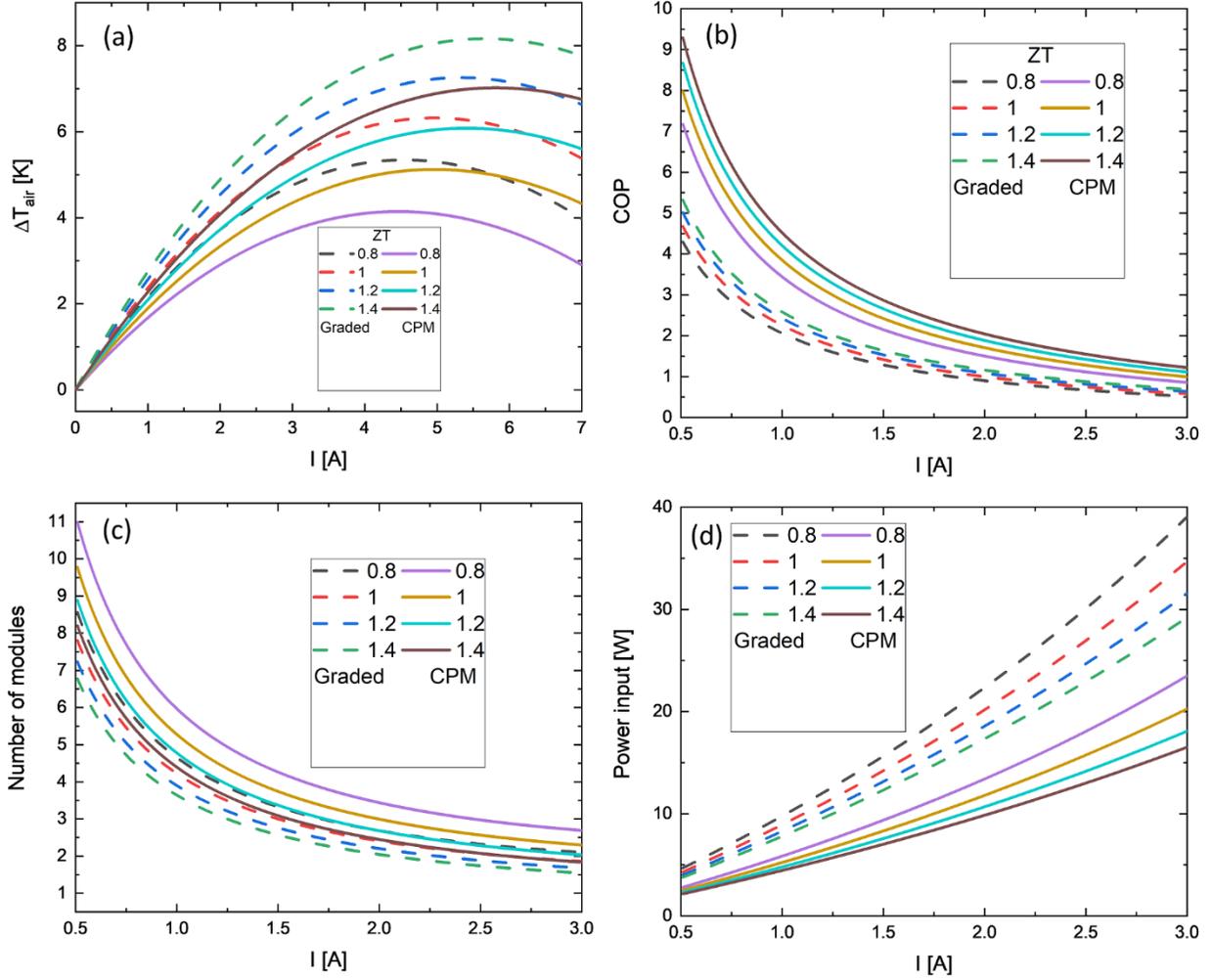

**FIG. 7** (a) Calculated degree of cooling per module, and (b) *COP*, (c) the number of modules required to achieve total $\Delta T_{air,target}$=10 K, and (d) total electric power consumption for one series of modules with the number of modules shown in (c) for different ZT values and for both CPM and graded segments (GPM). The fill factor and the leg thickness were fixed at $F = 0.1$ and $L = 2$ mm. Each leg cross-sectional area is 2 x 2 mm$^2$ and each module area is 5 x 5 cm$^2$. The air and water flow parameters used are given in Table 2. The effect of grading is not as significant on cooling especially at lower currents, and is about 2 K at currents higher than 2 A. This is consistent with the maximization of cooling by grading the TE leg. The *COP* is reduced to about half as compared to CPM, therefore, lesser power consumption in CPM. Meanwhile, for higher ZT, both cooling and *COP* are improved in both CPM and graded materials. The power consumption is only for TE device and excludes those for pumps and accessories.

## 4.3. Effects of air mass flow rate

In the previous sections, the air mass flow rate through a single series of channels was set to 0.002 kg/s which can be quite low compared to conventional air conditioning systems. So, in this section, effects of increasing air mass flow rates, to resemble more closely that of conventional A/C flow rates, on the device performance are investigated. To get a certain amount of air going into the system, we can use a number of modules thermally in parallel that are transverse to the direction of air flow. Mass flow rate is varied in two ways. First, the total mass flow rate is fixed, so the number of parallel modules will be inversely proportional to the mass flow rate through one series of modules, i.e. $N_{pm} = \frac{\dot{m}_{a,tot}}{\dot{m}_a}$. Hence, total mass flow rate $\dot{m}_{a,tot}$ is fixed to 0.05 kg/s first, and the device performance parameters are plotted with varying mass flow rate $\dot{m}_a$ through one series in Fig. 8. Electrical currents of 1 and 2 A are used for comparison, and both graded and CPM models are used ($ZT$ is kept constant at 1). Air cooling significantly drops with flow rate (Fig. 8(a)) because air moves along too fast before it is significantly cooled given a nearly constant heat flux by the TE. Its trends with the two material models and current are as expected and explained in Sec. 4.2. But since the same TE module is used, *COP* does not change much with flow rate (Fig. 8(b)). Note that there is a slight reduction in *COP* at low mass flow rates, simply because the constant currents used (1 and 2 A) are larger than the optimal current level in such low flowrates, so that Joule heating was excess. Due to the decreasing degree of cooling, the number of modules in series required to achieve a target temperature drop increases almost linearly with flowrate as can be seen in Fig. 8(c). However, the number of parallel modules decreases inversely proportional to flowrate, and therefore total number of modules (which is the product of the number of parallel modules and the number of modules in series), is roughly constant. This also results in the power input (Eq. (32)) being not so affected by the flow rate (Fig. 8(d)).

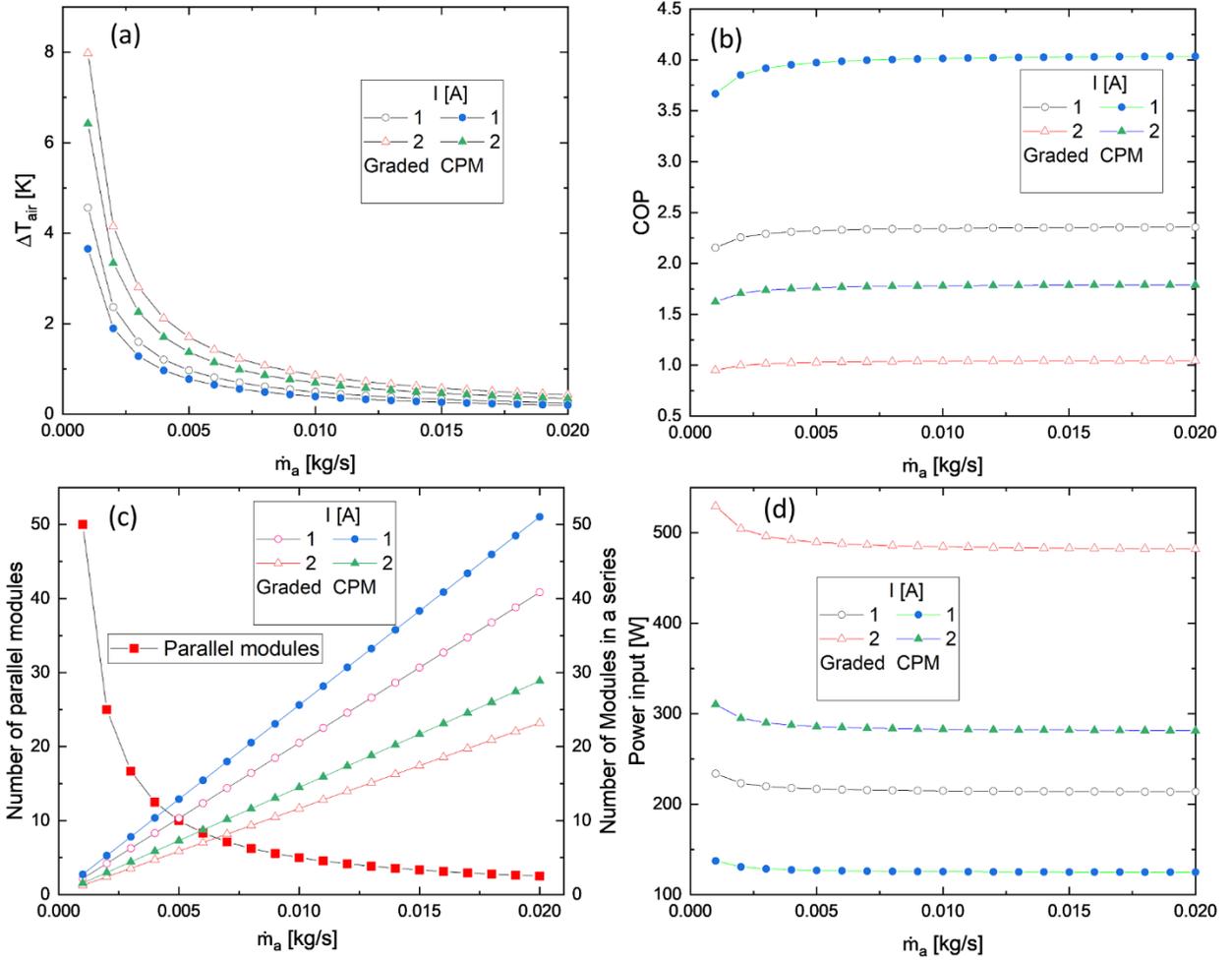

**FIG. 8** (a) Calculated degree of cooling per module, (b) *COP*, (c) the number of modules in series required to achieve total $\Delta T_{air,target}$=10 K, and number of parallel modules required for a total mass flow rate of 0.05 kg/s and (d) total electric power consumption for the whole system (all parallel and series modules); all varying with the air mass flow rate through a single series of modules ($\dot{m}_a$) at $ZT=1$ for both CPM and graded segments (GPM). Air cooling is affected drastically because of higher flow rates, but *COP* and power consumption are not. The number of modules in a series increases because of decreasing cooling per module, but the number of parallel modules decreases at the same time for a fixed total flow rate. Hence the total number of modules remains roughly constant, resulting in the mostly constant electrical power.

Secondly, the mass flow rate per module ($\dot{m}_a$) is changed to vary the total mass flow rate ($\dot{m}_{a,tot}$) while keeping the number of parallel modules constant at $N_{pm}=10$. The number of series modules has been adjusted to keep the total $\Delta T_{air,target}=13$ K as the increased mass flow rate per module results in a decrease in $\Delta T_{air}$ per module. The calculation results are displayed in Fig. 9. As shown in Fig. 9(a), the electrical power consumption increases linearly with total flow rate while the *COP* increases at very small flow rates and after 0.05 kg/s it becomes almost constant. Also shown in Fig. 9(a) is the comparison with three part-load curves for a good, typical and poor A/C defined in Ref. [36]. Full load values for these A/Cs are assumed to be: cooling capacity as 12000 BTU/h (1 Ton or 3.5 kW), energy efficiency ratio (EER) as 12 (COP=3.517) and air mass flow rate as 0.27 kg/s. To make a good comparison, cooling capacity is kept the same for both the TE and A/C at full load or at $\dot{m}_{a,tot}=0.27$ kg/s. This is the reason why $\Delta T_{air,target}$ is changed to 13 K as compared to 10 K in previous calculations. Part load power and *COP* are calculated based on the part load curve data for the three kinds of A/Cs provided by Henderson et al. [36] in their DOE report. In Fig. 9(a), TE cooler with the CPM model and 1 A current shows slightly lower power input than any of these A/Cs for the entire range of total mass flow rate because it has a higher *COP* than any other combination shown in the image. While the power input in graded model at 1 A is very close to the A/C data at lower flow rates, it gets higher at larger flow rates (~0.05 kg/s). The 2 A curves for both CPM and Grade model have higher power consumption in most of the flow rate range. Therefore, we find that low mass flow rate and low current are desired for the TE cooler to have a good *COP* and lower power consumption.

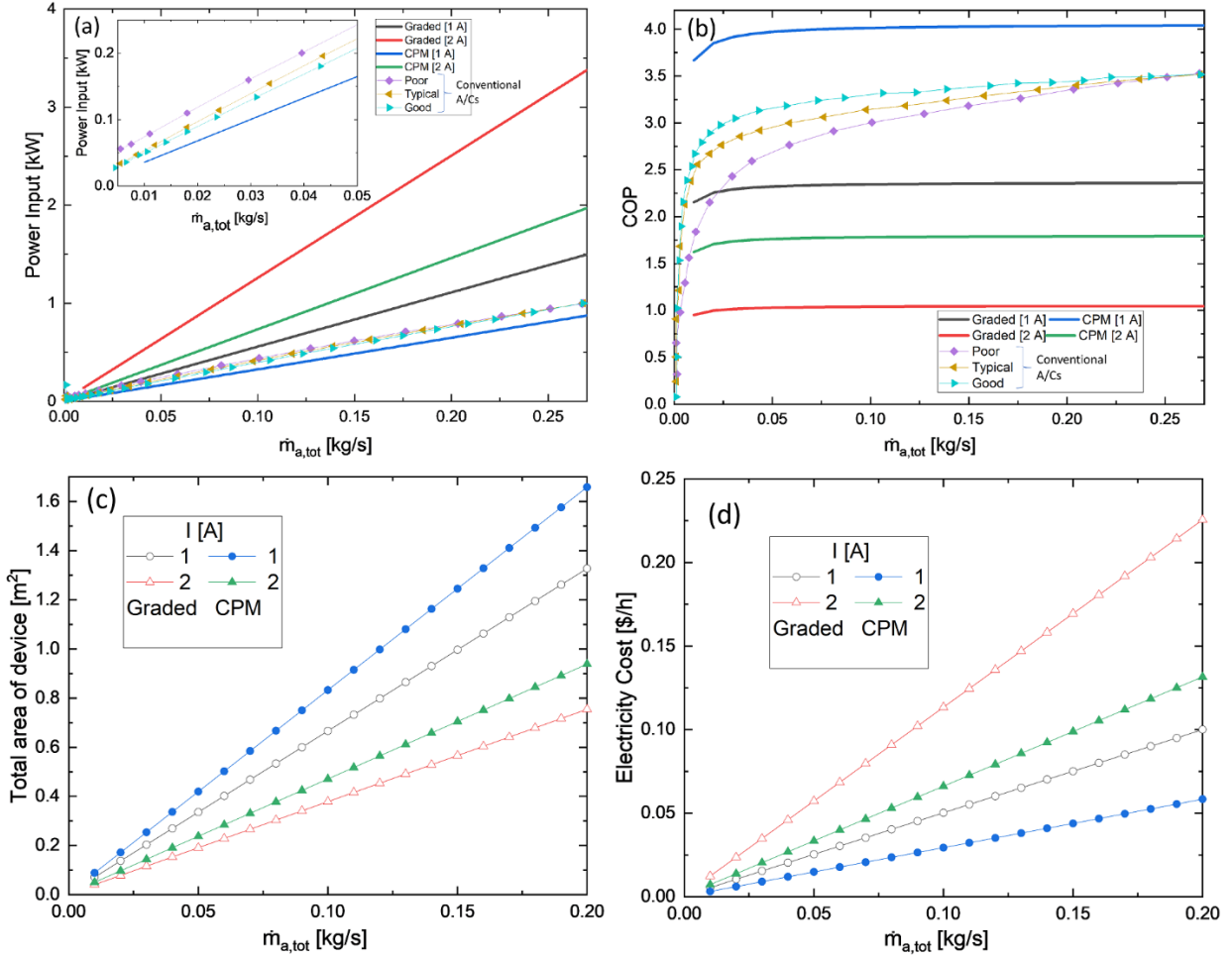

**FIG. 9** (a) Power input and (b) *COP* for TE coolers as a function of total air mass flow rate ($\dot{m}_{a,tot}$), in comparison with three typical residential A/Cs [36]; inset in (a) shows zoom-in plot at low flow rates up to 0.05 kg/s for CPM at 1 A with A/Cs. (c) Total top surface area of the device (($N_{pm} \times N_m \times 5 \times 5$ cm$^2$) and (d) electricity cost per hour for the TE coolers as a function of total mass flow rate of air at a constant *ZT*=1 for both CPM and graded segments (GPM) under 1 A and 2 A currents. Increasing total mass flow rate requires a larger number of modules in series ($N_m$) to keep total $\Delta T_{air,target}$=13 K with the constant number of parallel modules $N_{pm} = 10$ because the degree of cooling per module is reduced. This leads to higher area of the device, power consumption and hence the electricity cost for a constant current.

As is shown in Fig. 9(b), *COP* becomes fairly constant above 0.05 kg/s because the cooling capacity increases almost at the same rate as that of the power consumption with increasing mass flow rate. Total cooling capacity is independent of current and the material property models since the desired cooling is set to constant 13 K and the flow parameters do not change. Since cooling capacity at peak load is same for both TE cooler and the A/Cs, higher COP for the CPM model at 1 A leads to lower power consumption as shown in Fig. 9(a). Also, in terms of the part load *COP*, TE cooler having CPM at 1 A shows much better performance compared to the typical A/Cs, for which *COP* drops significantly as compared to the TE cooler. The CPM at 2 A and the graded models at 1 or 2 A showed lower COP than those of typical A/Cs over most of the mass flow rate range except for very low mass flow rates.

Fig. 9(c) displays the total top surface area of the device which is obtained by multiplying the module area ($5 \times 5$ cm$^2$) with the total number of required modules ($N_{pm}N_m$) as a function of total air mass flow rate. Total number of modules increases with increasing total air mass flow rate because higher total air mass flow rate causes cooling to be reduced per module, thus requiring a larger number of modules in series to achieve the target cooling of 13 K. This would require a longer duct and might increase the pressure drop as well. This also causes the area of the device to be higher. Based on Fig. 8 and Fig. 9, we find that keeping the total mass flow rate lower (0.05-0.2 kg/s) and low current operation (close to 1 A) would be more economical. Furthermore, CPM does better than the graded model based on the above results in terms of power consumption because of the higher *COP*.

### 4.4. Electrical power consumption

As shown in Fig. 9(a), TE coolers with CPM at 1 A have slightly lower power consumption than a typical 1 Ton A/C (about 125 W lower). This might get close to that of conventional A/Cs if the fan and pump power are included. Power consumption is higher for graded segments due to their much lower *COP* despite the reduced number of modules in series with a larger $\Delta T_{air}$ per module. Note that in Fig. 8 and 9 we used a *ZT*=1, which is the value for the state-of-the-art room-temperature TE materials. As we learned from Fig. 7(d), power consumption could be much reduced with a higher *ZT* value for both CPM and graded cases due to the enhanced Peltier cooling at the same current level. However, our results show that TE coolers even with the existing materials with *ZT* = 1 can achieve better cost effectiveness than conventional A/Cs. Design optimization with cost-performance trade-off in consideration is the key to the success of TE air conditioning at the market level. Since the cost is directly proportional to the power input, it has similar trends in Fig. 9(d) as for the power of our TE cooler in Fig. 9(a). The costs of CPM at 1 A will be very close to the A/C running cost because of the similarity in power consumption.

When the parameters are tuned to achieve better COP and lower power, cooling is reduced, needing higher number of modules and hence a larger system. This would lead to a lower running cost at the expense of a higher initial cost. This is a limitation for the thermoelectric technology when applied to cooling hence the above trade-off must be kept in mind while designing such a cooling system. In summary, low current operation (<1 A) and mass flow rate (~0.05 kg/s) can help in maintaining low power and a smaller device.

### 5. CONCLUSIONS

In this paper, we presented detailed modeling for thermoelectric air conditioning with both side convective heat exchangers, and provided strategies for cost-effective design of TE air coolers. We employed a 1D finite element method for the modeling of graded TE legs in the TE cooler. From our performance calculations, it is shown that the graded materials can produce better cooling than the modules with constant properties of the same figure of merit but at a significantly lower *COP*. Due to the reduced *COP*, power consumption is increased too much with the graded materials, which may not be suitable for cost-effective air conditioning. Our strategy for cost-effective design is to use a low current and low air mass flow rate with constant material properties to maintain a high *COP* and low power consumption as detailed in Section 4.3 and 4.4. This would mean that TE air conditioning is well suited for small space cooling dealing with low air flowrates. Even for large spaces, the total power input can still be comparable to or lower than those of typical A/Cs (Fig. 9(a)) with existing materials of *ZT* ~ 1, and flat-plate air-side heat exchangers. Improving the ZT value as well as enhancing the heat exchangers using advanced fin geometries will further lower the power consumption. Demand-flexible operation, small form factor, low noise, and modular structure, TE air conditioning can be very useful for a broad range of air cooling applications.

## ACKNOWLEDGMENTS

The authors thank Milind Jog and Raj Manglik for their helpful discussions on heat transfer analysis and comparison with conventional air conditioners. This material is based upon work supported by the U.S. Department of Energy, Office of Science, Office of Basic Energy Sciences Early Career Research Program under Award Number DE-SC0020154 (S.J.W.). J.-H. B. acknowledges support from NSF under Grant No. DMR-1905571.

**Nomenclature**

| | | | |
|---|---|---|---|
| $A$ | cross-section area of air channel (cm$^2$) | $N_m$ | number of modules along the direction of air flow |
| $A_l$ | TE leg area (mm$^2$) | $N_{pm}$ | number of modules transverse to the direction of air flow |
| $A_s$ | surface area of air heat exchanger (cm$^2$) | $n_s$ | number of air channels in a heat exchanger |
| $A_{sw}$ | surface area of water duct (cm$^2$) | $Nu$ | nusselt number |
| $A_t$ | single module area (cm$^2$) | $P$ | perimeter of air channel (cm) |
| $c$ | doping ratio | $Pr$ | prandtl number |
| $COP$ | coefficient of performance of the TE device | $Q_c$ | cooling capacity of a TE module (W) |
| $C_p$ | specific heat capacity (J/kgK) | $Q_h$ | heat transfer on water side (W) |
| $d$ | density (kg/m$^3$) | $Q_J$ | joule heating (W) |
| $D$ | diameter of water duct (cm) | $Q_K$ | conduction heat (W) |
| $D_h$ | hydraulic diameter of air channel (cm) | $Q_P$ | peltier cooling (W) |
| $F$ | fill factor | $R$ | electrical resistance (Ω) |
| $H$ | height of air heat exchanger (cm) | $R_c$ | electrical contact resistance (Ω) |
| $h'$ | heat transfer coefficient (W/m$^2$K) | $Re$ | reynold's number |
| $h$ | effective heat transfer coefficient (W/m$^2$K) | $S$ | seebeck coefficient (μV/K) |
| $I$ | electric current (A) | $T$ | temperature (K) |
| $k$ | thermal conductivity (W/mK) | $V_s$ | seebeck voltage (V) |
| $k_g$ | gap filler thermal conductivity (W/mK) | $W$ | width of air heat exchanger (cm) |
| $K$ | thermal conductance (W/K) | $W_m$ | power input for a module (W or kW) |
| $L$ | depth of a module (cm) | $x$ | position along a graded TE leg length (mm) |
| $L_l$ | TE leg thickness (mm) | $ZT$ | thermoelectric figure of merit |
| $L'$ | length of water duct per module (cm) | | |
| $\dot{m}$ | mass flow rate (kg/s) | | |
| $n$ | number of nodes in a graded leg | | |
| $N$ | number of TE leg pairs in a module | | |

*Subscripts*

| | | | |
|---|---|---|---|
| $a$ | air | | |
| $c$ | cold side of TE module | | |
| $h$ | hot side of TE module | | |
| $i$ | index for a node in a graded TE leg | | |
| $ia, fa$ | initial and final for air | | |
| $iw, fw$ | initial and final for water | | |
| $n$ | n-type TE leg | | |
| $p$ | p-type TE leg | | |
| $tot$ | for the whole device | | |
| $w$ | water | | |

*Greek symbols*

| | |
|---|---|
| $\alpha$ | aspect ratio of air channel |
| $\Delta T_{air}$ | degree of cooling in air (K) |
| $\Delta T_{air,target}$ | target degree of cooling of air |
| $\mu$ | dynamic viscosity (Ns/m$^2$) |
| $\rho$ | resistivity (Ωm) |
| $\sigma$ | electrical conductivity ((Ωm)$^{-1}$) |

*Universal constants*

| | |
|---|---|
| $e$ | charge on electron |
| $k_B$ | Boltzmann constant |

*Abbreviations*

| | |
|---|---|
| A/C | Air Conditioner |
| CPM | Constant Property Model |
| GPM | Graded Property Model |
| TE | Thermoelectric |